\documentclass[letterpaper]{article}

\usepackage{natbib,alifeconf}  
\usepackage{url,hyperref,cleveref}
\usepackage{booktabs}
\usepackage{caption}
\usepackage{subcaption}

\title{Analysis of a Spatialized Brain-Body-Environment System}

\author{
    Denizhan Pak$^{1,2}$,
    Quan Le Thien$^{3,4}$,
    \and Christopher J. Agostino$^{5}$
   \mbox{}\\
    $^1$ Department of Cognitive Science, Indiana University, Bloomington, IN 47405, USA \\ 
    $^2$ Luddy School of Informatics, Computing, and Engineering, Indiana University, Bloomington, IN 47405, USA \\
    $^3$ Department of Physics, Indiana University, Bloomington, IN 47405, USA\\
    $^4$ Quantum Science and Engineering Center (QSEC), Indiana University, Bloomington, Indiana 47405 \\
    $^5$ NPC Worldwide, Bloomington, IN 47403, USA (info@npcworldwi.de)  
} 

\begin{document}

\maketitle

\begin{abstract}
The brain-body-environment framework studies adaptive behavior through embodied and situated agents, emphasizing interactions between brains, biomechanics, and environmental dynamics. However, many models often treat the brain as a network of coupled ordinary differential equations (ODEs), neglecting finer spatial properties which can not only increase model complexity but also constrain observable neural dynamics. To address this limitation, we propose a spatially extended approach using partial differential equations (PDEs) for both the brain and body. As a case study, we revisit a previously developed model of a child swinging, now incorporating spatial dynamics. By considering the spatio-temporal properties of the brain and body, we analyze how input location and propagation along a PDE influence behavior. This approach offers new insights into the role of spatial organization in adaptive behavior, bridging the gap between abstract neural models and the physical constraints of embodied systems. Our results highlight the importance of spatial dynamics in understanding brain-body-environment interactions.
\end{abstract}



\section{Introduction}
Adaptive behavior is often modeled using a brain-body-environment framework \citep{ChielBeer1997}. Within this framework, the body and environment are often modeled as a mechanical systems. Neural dynamics couple with this mechanical system as a collection of ordinary differential equations (ODEs), often implemented as a recurrent neural network \citep{Beer2003}. This approach has been foundational to the study of adaptive behavior and has yielded many experimentally validated insights \citep{Beer2000}. 

However, this paradigm faces some theoretical limitations in its ability to model some of the geometric relations that can shape neural dynamics. Spatial properties of neural networks are reduced to edge weights between nodes, obscuring the spatial extent of the nodes themselves and the complex geometry observed in real neural tissue \citep{PaikEtAl2020}. Additionally, while external time delays can be added to ODEs, PDEs provide a natural trade-off between space and time that demonstrates delay dynamics.

Neural field models, which emphasize spatial dynamics, have a long history in neuroscience \citep{CoombesBeimGrabenPotthastWright2014}. However, research integrating neural fields with biomechanical simulations remains underexplored \citep{dale_evolution_2010}. Partial differential equations (PDEs) are natural tools for neuromechanical modeling, as they capture both temporal and spatial dynamics critical to understanding adaptive behavior within brain-body-environment frameworks.

PDEs offer a different perspective from ODE-based neural mass network models \citep{PinotsisEtAl2013}. Their vector fields depend on both time and space, enabling them to incorporate spatial constraints imposed by embodied interactions with the environment. Additionally, PDEs can model signal propagation more accurately: while ODEs assume instantaneous signal transmission across nodes, PDEs explicitly represent spatial signal propagation through a medium. Therefore, they can be used to model delays in signaling, gradients of activity and neural wave patterns. The cost of this expressivity is increased complexity. However, simplifying assumptions—as we demonstrate—can mitigate this complexity and enable analytical tractability.

We explore PDE-based neuromechanical modeling through the lens of minimal cognition \citep{Beer1995}. Minimal cognitive modeling studies adaptive behavior in simplified brain-body-environment systems performing tasks that serve as conceptual analogs to real-world complexity \citep{Beer2003}. This approach has proven effective for clarifying philosophical interpretations \citep{VarelaThompsonRosch1991, Clark1997}, developing new tools for model creation and analysis \citep{Beer1995}, and inspiring experimental applications \citep{marder_theoretical_2020}.

For this work, we adopt the reactive swinging agent system introduced by Thorniley and Husbands (\citeyear{ThornileyHusbands2013}). This model simulates a child swinging on a swing, aligning with minimal cognitive modeling due to its simplicity in biomechanical and environmental components. The system also features well-defined optimal behavior within specific parameter ranges, facilitating the conversion of its original ODE-based neural controller into a PDE framework.

The remainder of this paper is structured as follows: Section 2 details Thorniley and Husbands’ original model, including a bifurcation analysis of the sensory-coupling parameter. Section 3 describes the reformulation of the agent’s state variables as PDEs, highlighting dynamical differences from the original model. Section 4 introduces a simplifying assumption to analyze these differences. Section 5 explores the implications of this assumption. Finally, Section 6 examines signal propagation through spatialized sensors and effectors, demonstrating how spatial parameters influence task performance.

\section{A Simple Brain-Body-Environment System}
\begin{figure}
    \centering
    \includegraphics[width=0.7\linewidth]{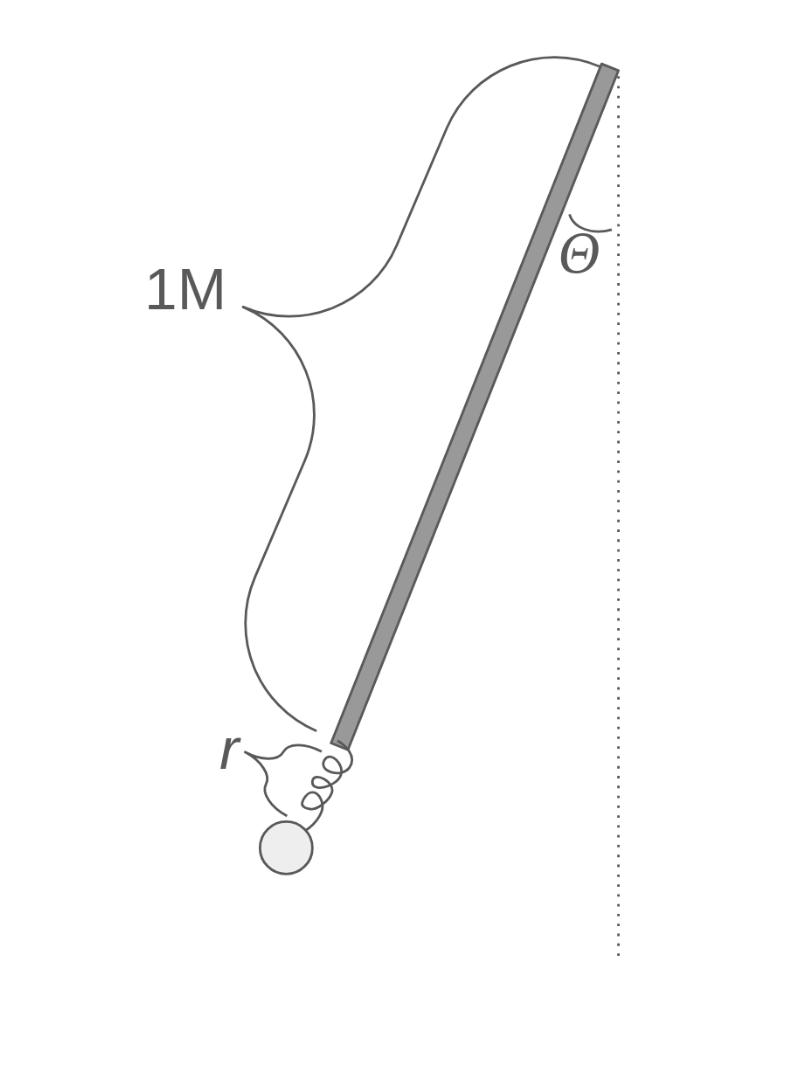}
    \caption{The neuromechanical model originally presented in \cite{ThornileyHusbands2013}. The agent is represented as a mass-spring which is attached to then end of a rigid rod. The rod can rotate about the top and $\theta$ denotes the deviation from the rod in its steady resting state. The dynamics of the movement of the rod are governed by the gravitational and torsional forces which are affected by the length of the agent $r$. The agent can then actuate the value of $r$ to move the rod (swing).}
    \label{fig:system}
\end{figure}
Our system of interest is the reactive swinging agent proposed by \cite{ThornileyHusbands2013}. The swing is modeled as a rigid, massless rod attached to a fixed pivot at one end, with a mass (representing the agent) at the other end. We visualize this system in figure \ref{fig:system}.

The original system is governed by the following equations:
\begin{eqnarray}
\dot{\theta} = \omega \\
\dot{\omega} = -\frac{g}{1+r}\sin(\theta) - b \omega \\
\label{r eqn}
\dot{r}= v \\
\dot{v} = \frac{u}{1+r^2} + g \cos(\theta)+(1+r)\omega^2 - kr -cv \\
\dot{u} = \phi(A \tanh(\rho v) - u) 
\end{eqnarray}

This 5-dimensional system of ordinary differential equations (ODEs) separates into environmental states $(\theta, \omega)$ and agent states $(r, v, u)$. Here, $\theta$ is the swing angle, $\omega$ the angular velocity, $r$ the pendulum extension controlled by the agent, $v$ the extension rate, and $u$ the neural force acting on $v,r$. Descriptions and numerical values can for those and other parameters can be found in Table \ref{tab:params}.

\begin{table*}
    \centering
    \begin{tabular}{ccc}
        Symbol & Type/(Initial Value/Range) & Description\\
        \hline
        $\theta$ & Variable ($\pi$) & Angle of pendulum from downward vertical\\
        $\omega$ & Variable ($0.1$) & Angular velocity of pendulum\\
         r& Variable ($0.1$)& Current pendulum extension\\
         v& Variable ($0.1$)& Rate of pendulum extension\\
         u& Variable ($0.1$)& State variable of agents nervous system\\
         $A$& Parameter $(0,300)$ & Strength of sensorimotor coupling\\
         g& 9.81 & Acceleration due to gravity\\
         b& 0.3 & Pendulum damping coefficient\\
         $\rho$& 2 & Motor neuron sensitivity\\
         $\phi$& 20 & Control parameter\\
         k& 100 & Spring force constant\\
         c& 20 & Spring damping (= $2\sqrt{k}$ for critical damping)\\
         $x_s$ & Parameter $(-1,1)$ & Central spatial position on $v$ that is driven by $r$\\
         $\sigma_s$ & Parameter $(0,\infty)$ & Standard deviation of the spatial part on $v$ that is driven by $r$\\
         $x_e$ & Parameter (-1,1) & Spatial location $v(x_e,t)$ that drives by $r$\\
         $(x_{\sf min},x_{\sf max})$ & Parameter $(-1,1)$ & Boundary points of $v$ and $u$
    \end{tabular}
    \caption{Table of variables and parameters, adapted from \cite{ThornileyHusbands2013}. The numbers next to the variables represent their initial value used for the simulations. The numbers next to parameters represent their ranges across plots.}
    \label{tab:params}
\end{table*}

To analyze the system, we varied the parameter $A$, which governs the strength of sensory coupling strength between $u$ and $v$. Previously, \cite{ThornileyHusbands2013} identified three dynamical regimes as $A$ increases:

\begin{enumerate}
    \item \textbf{Dampened regime:} In this regime, the system dampens to a fixed point neither the environment nor the agent show any interesting long-term behavior. This can be thought of as the swing stopping.
    \item \textbf{Oscillatory regime:} In this regime, the system converges to a fixed amplitude and period oscillation with a bounded $\theta$. This regime can be thought of as successful swinging in which the swing goes back and forth.
    \item \textbf{Chaotic regime:} In this regime, the system displays a broad range of oscillations of various amplitude and frequency sometimes moving fast sometimes slowly. This can be thought of as the regime where the swing loops around the top, moving too fast to control.
\end{enumerate}

\begin{figure}
    \centering
    \includegraphics[width=1\linewidth]{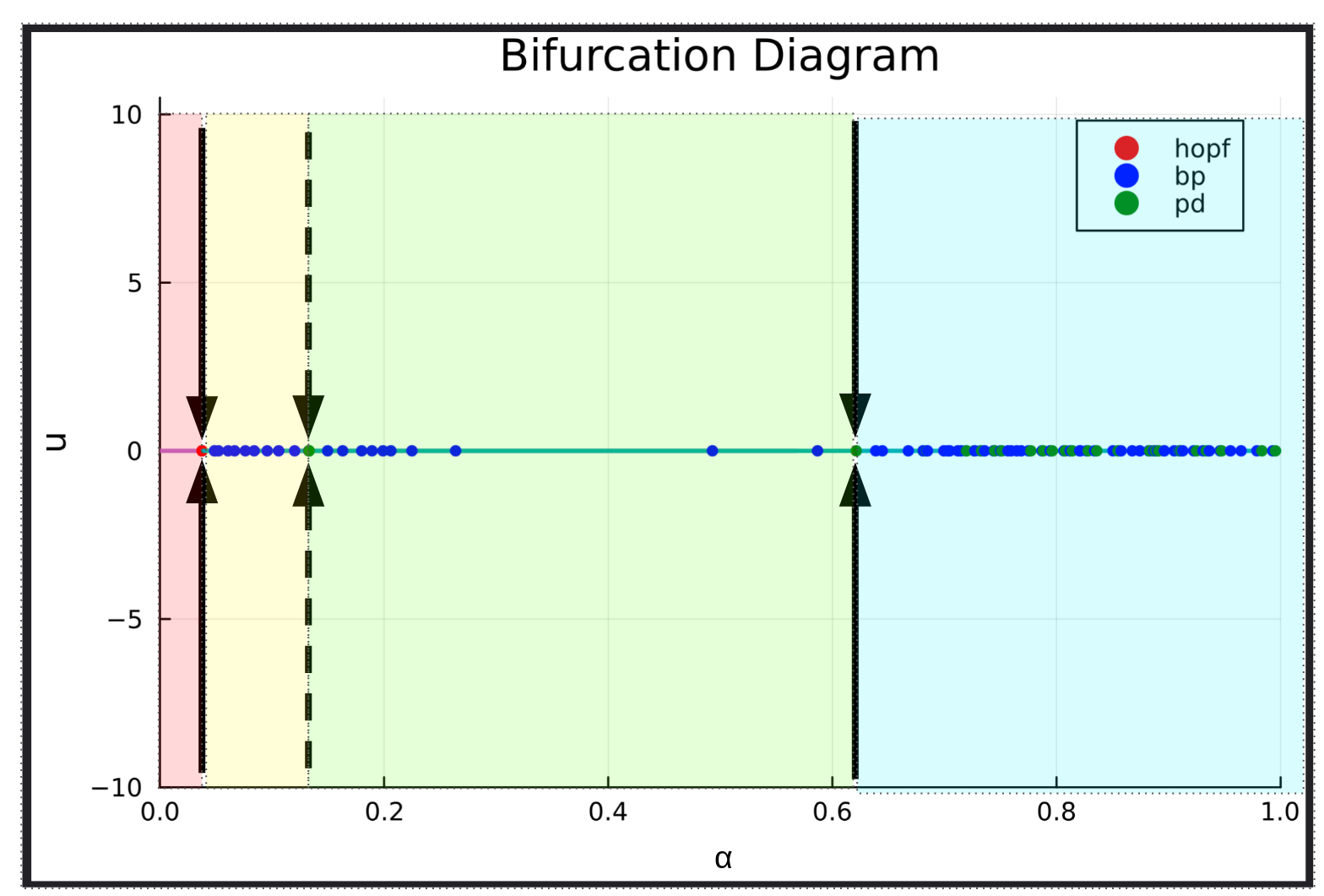}
    \caption{Bifurcation diagram for a swinging agent system, plotting agent state $u$ (vertical axis) at the vertical stable position against sensory sensitivity $\alpha$ (horizontal axis). We used a normalized parameter $\alpha=\frac{A}{300}$ for computational simplicity. $u$ remains stable at zero until $A$ exceeds a threshold, causing the swing to overturn. Key transitions include the Hopf point (onset of oscillations), pd (period-doubling cascade into chaos), and bp (branch points for computational continuation). The coloring and solid arrows in the bifurcation diagram represent the 3 phases of the system: Dampened regime (red), oscillatory regime (yellow), chaotic regime (blue). The diagram also shows that the transition from oscillatory to chaotic is not instantaneous as it requires a cascade of pd bifurcations, the transitory region is tinted green and denoted with a dashed arrow.}
    \label{fig:bif_diag}
\end{figure}

To quantify the onset and better understand the nature of these regimes we performed a bifurcation analysis. This analysis was done using the BifurcationKit library from the Julia programming language. We first found the one and only equilibrium point in the dampened regime. Having identified this point, we performed continuation along the $A$ parameter which lead to a Hopf bifurcation. This explains the onset of the oscillatory regime \citep{Kuznetsov2006Hopf}. This allows for a clear demarcation point between the onset of oscillatory and the dampened behaviors. However, quantifying the onset of the chaotic regime was slightly more difficult. After the first round of continuation, we found several branch points along the same initial branch. By performing continuation along each of these branch points individually, we identified one of these branch points as a period-doubling bifurcation. Continuation along the period-doubling bifurcation showed that it was indeed the onset of chaos \citep{Meiss2007PeriodDoubling}. This implies that the mechanism for chaos in this system is the period-doubling route, a common universality class in a wide range of dynamical systems \citep{Strogatz2018}. We plot the results of our analysis as a bifurcation diagram in figure \ref{fig:bif_diag}.

Thus our results demonstrate that the same three regimes discovered in the original model have a standard dynamical basis. Namely, that there is a Hopf bifurcation that separates the dampened and oscillatory regime and that the oscillation resulting from this initial Hopf bifurcation takes a period-doubling route to chaos that leads to formation of a chaotic regime. We also note that the initial period-doubling coincides with the formation of a second limit cycle. This is when the swing is stable not only going back and forth but also swinging around the top.

\section{Spatializing the Brain and Body}
\begin{figure*}
    \centering
    \includegraphics[width=.7\linewidth]{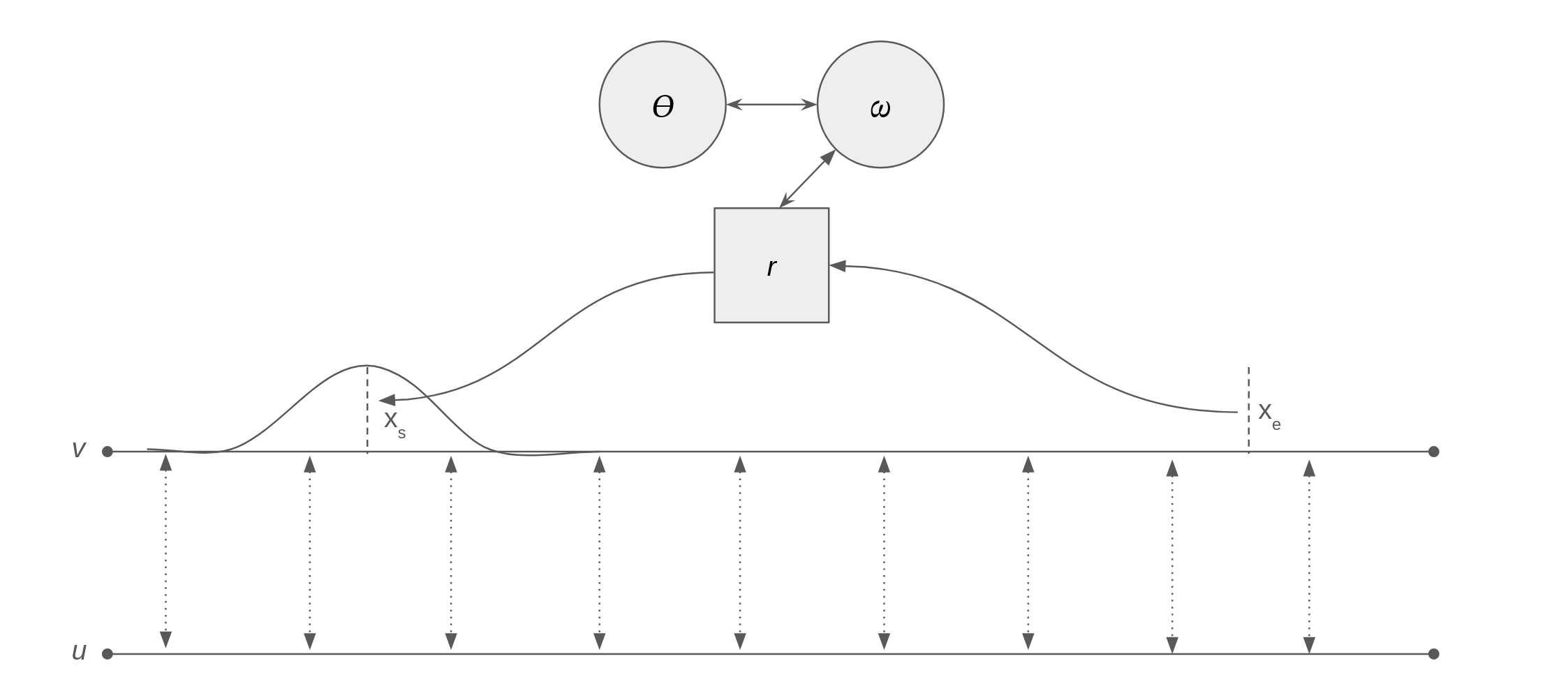}
    \caption{Here we visualize the variables of the agent as a circuit diagram. $\theta$, $\omega$ and $r$ are three state variables. Variables in a circle represent underactuated environmental variables. The variable $r$ is the barrier between the agent and the environment, playing both roles as the agent's sensor, gathering information from $\omega$ into the weighted region $p(x-x_s)$ centered at $x_s$ in the agent $v$, and as the agent's actuator, realizing the agent's response from $v(x_e,t)$ into the environment $\omega$. Note that we assume the distribution $p(x-x_s)$ to follow $N(x_s,\sigma_s)$, we examine both cases $\sigma_s\rightarrow\infty$ (uniform distribution) and finite $\sigma_s$ (localized distribution). Points on the ends of the state variables $x_{\sf min}$ and $x_{\sf max}$ denote where the Dirichlet boundary conditions are specified.}
    \label{fig:agent-diagram}
\end{figure*}

We modify the system by promoting $u(t)\rightarrow u(x,t)$ and $v(t)\rightarrow v(x,t)$. This converts these two originally one-dimensional ODEs into PDEs, while maintaining $\dot{r},\dot{\omega},\dot{\theta}$ as ODEs. The resulting system becomes infinite-dimensional due to the spatial extension of $\dot{u}$ and $\dot{v}$, while remaining a mixed PDE-ODE system.

We note there are multiple approaches to generalizing ODEs to PDEs, depending on the interpretation of the spatial variable $x$. For instance, $x$ could represent a state variable where $v(x,t),u(x,t)$ denotes the probability of observing $x$ at time $t$ - in such cases, one might employ a Fokker-Planck equation \citep{Risken1996}. However, we adopt a different approach here by interpreting $x$ as representing position along a one-dimensional neural tissue.

While real neural tissue is fundamentally three-dimensional (though sometimes modeled as two-dimensional \citep{HuttRougier2010}), our present goal is to demonstrate the basic effects of spatializing neural dynamics in embodied agents, rather than to model specific neuroanatomical structures. We therefore consider $u$ and $v$ to represent neural activity along two separate one-dimensional neural tracts, which provides sufficient complexity to examine our core questions while maintaining analytical tractability.

Conceptually, our spatialization of the $u,v$ system assumes not at a single point (as implied by the ODE formulation) but rather along a continuous one-dimensional neural tissue. Similarly, the dynamics of $v$, which represent the agent's body, also extend along an axis parallel to the nervous system. The variable $r$, representing the agent's length, naturally remains a one-dimensional quantity. $\theta$ and $\omega$ could theoretically be extended to PDEs, our physical assumptions - a rigid, massless rod fixed at the base - imply these quantities remain uniform along the rod's length, justifying their treatment as one-dimensional ODEs.

The introduction of spatial dimensions to $u$ and $v$ raises important questions about intercellular coupling. Various coupling schemes exist in the neural field literature \citep{CoombesBeimGrabenPotthastWright2014}. For our minimal model, we adopt the simplest case of lateral coupling: each infinitesimal segment (mathematical point) of $u$ and $v$ interacts with its neighbors through a distance-dependent coupling. This corresponds to a diffusive interaction, which can alternatively be interpreted as a rate-based approximation of discrete spike-based signaling \citep{FaugerasEtAl2009}. Formally, we add diffusion terms to the original $u$ and $v$ equations over the spatial dimension, following established techniques in continuum neural modeling \citep{Bressloff2012}. The resulting equation then becomes:
\begin{eqnarray}
    \partial_t{v} &=&
K \partial_{xx} v + \frac{u}{1+r^2}  -cv \\
\nonumber && +p(x-x_s) \left[ g \cos(\theta)+(1+r)\omega^2 - kr \right]\\
\partial_t{u} &= & K \partial_{xx} u+ \phi(A \tanh(\rho v) - u)
\end{eqnarray}
where the introduction of diffusion brings a new parameter $K$ representing diffusivity. Meanwhile, $p(x-x_s)$ follows the spatial distribution $N(x_s,\sigma_s)$ and screens out the spatial portion in the agent system that is coupled directly to the environment. We first consider the uniform global coupling case where $\sigma_s\rightarrow\infty$. 

Since the agent state $v(x,t)$ is now a PDE variable, we need to specified the point $x_e$ from which the actuator $r$ takes in the agent's output $v(x_e,t)$ and realizes the agent's decision into the environment. This means that Eq.~(\ref{r eqn}) is now promoted to
\begin{eqnarray}
    \dot{r}= v(x_e,t) 
\end{eqnarray}
We fix $x_e=0$ at the center of the string for all simulations. The distance $x_s - x_e$ signifies how far the sensory signal around $x_s$ has to diffuse through the agent to its actuators while being processed.

Before examining the effects of this distance, we must address several modeling assumptions regarding system couplings:
\begin{enumerate}
    \item \textbf{Coupling between $r$ and $v$}:
We spatialize the input from $r$ to $v$ as a Gaussian centered at a point $x_s$, which for now we take to be the midpoint of the string $v$. Since $v$ is now continuous, $r$ must couple to a region rather than a single point - equivalent to assuming sensory signals are received by a Gaussian-distributed patch of neural tissue. Similarly, we must define an effector point $x_e$ which is the point at which the effects of $v$ affect $r$. Which for now we take to be the midpoint as well.
\item \textbf{Coupling between $u$ and $v$}:
Drawing biological inspiration from nerve cords in advanced mammals, we model $u$ not as a centralized brain but as distributed processing interacting locally with the body $v$. Mathematically, this means $u$ and $v$ are coupled point-wise along their spatial extent.
\item \textbf{Boundary conditions}:
We implement Dirichlet boundary conditions 
\begin{eqnarray}
    u(x_{\sf min},t)&=&BC(x_{\sf min})\\
    u(x_{\sf max},t)&=&BC(x_{\sf max})\\
    v(x_{\sf min},t)&=&BC(x_{\sf min})\\
    v(x_{\sf max},t)&=&BC(x_{\sf max})
\end{eqnarray}
where $BC= D (x-x_{\sf min})$ with D representing the bias slope. At first, we will examine the case $D=0$ and then move on to the bias model where one end of the agent is fixed at higher activity.
\end{enumerate}

We diagram these assumptions in firgure \ref{fig:agent-diagram}.

The diffusion parameter $K$ introduces non-trivial effects despite its linear formulation, due to interactions with nonlinear terms and the boundary conditions. To better understand the implications of a spatial formulation we tried to unravel the effects of $K$ in relation to the three regimes we described in the previous section. However, some crucial difficulties emerge when trying to understand how bifurcations in ODEs are related to the behavioral regimes of PDEs.

The first distinction that we highlight is the dynamics of the new system are not spatially uniform. In the case where $K = 0$ (and $x_s=x_e =x_0$), the dynamics of the PDE system for the point $x_0$ is equivalent to the original ODE formulation. This is because when $K=0$ there is no diffusion and since the effector and sensor are assumed to be at the same point, the dynamics occurring at $x_0$ are simply the same as the point mass assumption of the original ODE system. Points within the Gaussian input from $r$ also display similar dynamics, but the amplitude of their behavior is reduced proportional to their distance from $x_s$. This can be understood as the original system still being localized to $x_0$ since $K=0$ implies no lateral coupling but with the sensory signals from $r$ to $v$ broadcasting across $x$ according to our Gaussian assumption.

As we increase $K$, it introduces diffusion between neural elements. This means that $x_0$ must also now interact with its neighbors. The result of this interaction is that the original system behavior, again localized at $x_0$, is now influenced by the spatial extant of the $u,v$ system. However, since $K$ is a diffusion term, it causes the excitation in $x_0$ to dissipate away into the rest of the string. In the next section, we will explain how this relates to our choice of boundary condition. For now, we take this mean that the contribution of $K$ primarily is to dissipate the energy generated in the coupling between the agent and the environment into the spatial dimension of the agent. This implies that at higher values of $K$ we should observe that a larger value of $A$ is required to observe the onset of the non-equilibrium regimes.

From this perspective, it would be intuitive to suggest that the role of $K$ is to shift the locus of the bifurcations relative to the $A$ parameter values that we observed in figure \ref{fig:bif_diag}. However, we note another important complication. Bifurcations are relatively well-defined in the context of ODEs but in the case of PDEs this is not the case. Since the dynamics are infinite dimensional, it is hard to define a single point at which the system shifts behavior qualitatively since it may shift along some points in $x$ but not others. Instead, PDEs are known to exhibit a phenomena known as criticality which is different than bifurcations. A detailed description of the differences between bifurcations and criticality is outside the scope of this paper, we direct to \cite{bose_bifurcation_2019} as a useful starting point. For our purposes, the distinction that matters is that bifurcations can be defined simply in terms of a parameterized dynamical system, whereas criticality requires a system taken together with an observable (i.e. a function of $x$) that can be used to summarize the behavior of the system as a whole. Criticality is the qualitative change in this observable rather than in the fundamental system.

\begin{figure}
    \centering
    \includegraphics[width=1.1\linewidth]{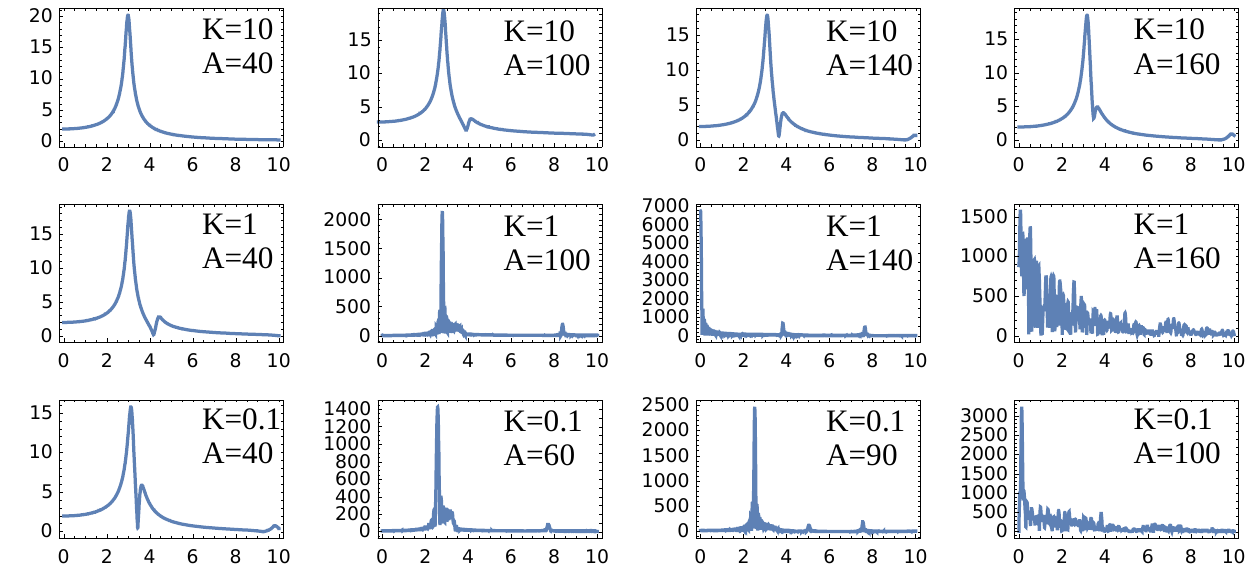}
    \caption{Frequency power spectrum of the angular velocity $\omega(t)$ as the diffusivity $K$ and the sensory coupling strength $A$ are varied. The effector position is fixed at $x_e=0$ with global presence in the agent $\sigma_s\rightarrow\infty$. The boundary condition is $D=0$, and thus $BC(x)=0$. We observe a variety of phases of the mixed PDE-ODE system in terms of parameters $K$ and $A$. As $A$ increases we see the sequence of transitions from the dampened regime to the oscillatory regime to the chaotic regime. As $K$ increases this process is delayed prolonging the dynamics of the stable regimes even with high sensory feedback.}
    \label{fig:crit-diag}
\end{figure}

Having familiarized ourselves with the dynamics of the system, we considered a simple observable to be the Fourier transform. A Fourier transform is used to identify the underlying periodic behaviors that make up the dynamics of the system. We direct readers to \cite{lange_fourier_2020} for a clear introduction. The Fourier spectrum would help us identify which of the three phases was dominating the system. In the dampened regime we would expand a band of low frequency oscillations that would have small amplitude. In the oscillatory regime there should be a singular dominant frequency. In the chaotic regime, we would expect a range of different frequency with different amplitudes. We plotted this observable as a function of the $K$ and $A$ parameters in \ref{fig:crit-diag}.

The resulting phase diagram shows many of the same features as the previous bifurcation diagram. We see again that $A$ acts as a sort of temperature, as it increases the system moves towards the more active regimes where eventually the temperature reaches a second critical point and at that point we see the onset of the chaotic regime equivalent in the PDE. We can see that the role of $A$ is mitigated by $K$ which acts a heat capacity making the system require a greater $A$ to overcome the diffusion between the points along the neural string.

\section{Approximating Signal Diffusion with a Mental Bias}
As demonstrated above, spatial diffusion—controlled by the parameter $K$—regulates how sensory input propagates through the system. This underscores why sensory systems should not be modeled merely as discrete interaction points between an agent and its environment. Instead, they are better represented as continuous surfaces that not only respond to perturbations but also maintain intrinsic stability. However, the current formulation of $K$ is somewhat arbitrary, representing a global parameter that lacks biological nuance. Moreover, simulating the full coupled PDE-ODE system to study parameter effects is computationally expensive. Fortunately, PDEs—widely studied in physics—offer approximation techniques to simplify such analyses. Here, we focus on the boundary distance approximation.

Before we elaborate on this approximation, we must first discuss, in more detail, the role of boundary conditions. PDEs are defined along a continuous spatial extant. In theory this extant could be infinitely long, however, for any computational simulation we assume some finite length for the spatial dimensions of the PDE ($x_{min},x_{max}$). This range allows us to simulate the PDE on a computer. It also makes sense that the physical system we are modeling is not infinitely large. However, this introduces a problem that the dynamics are not defined beyond $x_{min}$ and $x_{max}$. To simulate the system then we must introduce some explicit constraints that specify how the simulation will treat values of $x$ that in theory would interact with points beyond $x_{min}$ and $x_{max}$. The simplest treatment is to fix those points at a specific value, in our previous simulation this was the Dirichlet boundary condition where those values where set to $0$.

The boundary conditions can have complicated effects when dynamics are non-linear but when dynamics are purely diffusive and linear they act as attractors. This means that dynamics near the boundaries converge to whatever value the points at $x_{min},x_{max}$ are set to be. The role of $K$ can also be understood as modulating the strength of this attractor dynamic. This linear story is incomplete since our dynamics involve a non-linear feedback. However, this approximation was suggestive enough to warrant further investigation.

\begin{figure}
    \centering
    \includegraphics[width=1.1\linewidth]{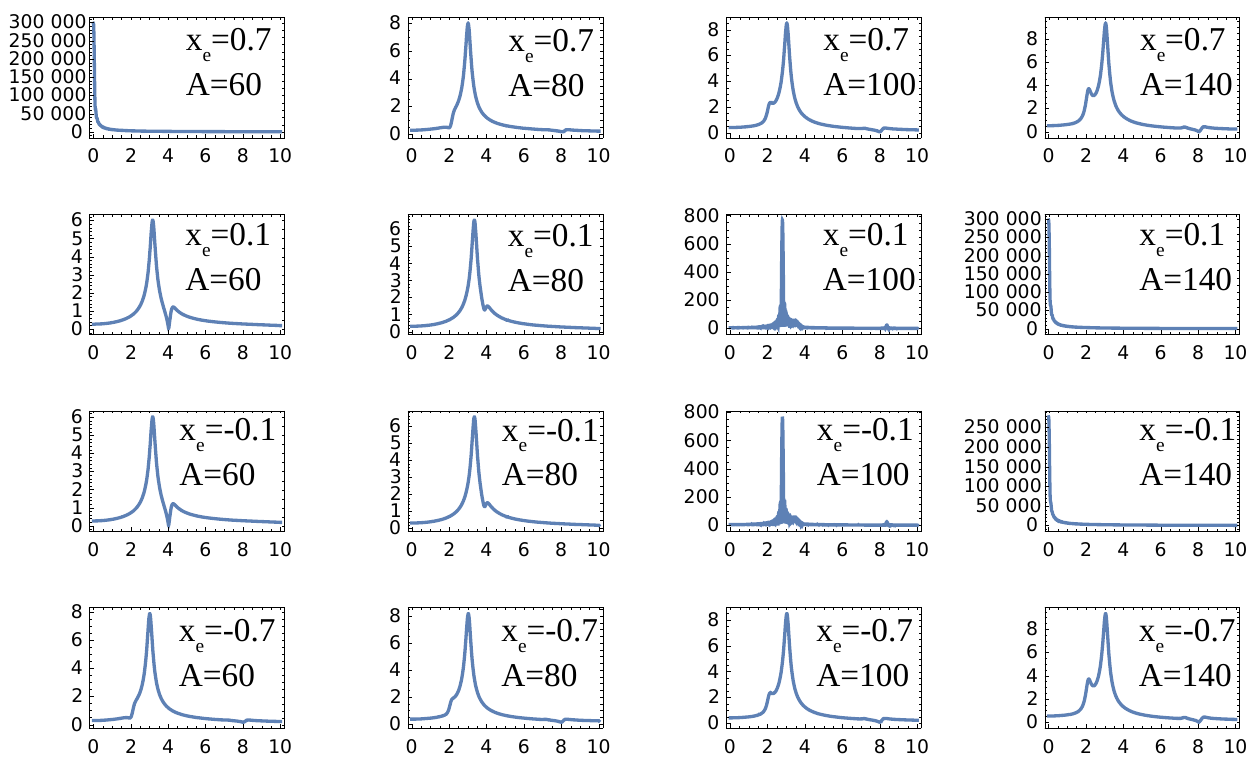}
    \caption{Frequency power spectrum of the angular velocity $\omega(t)$ as the effector position $x_e$ and the sensory coupling strength $A$ are varied. The diffusivity is fixed at $K=1$ and the envionment interacts globally in the agent $\sigma_s\rightarrow\infty$. The boundary condition is $D=0$, and thus $BC(x)=0$. We observe a similar variety of phases of the mixed PDE-ODE system but the transition point and the approach to criticality is modified due to the influence of the Dirichlet boundary condition on the effector is stronger when the effector position $x_e$ is closer to the two ends.}
    \label{fig:xe-diag}
\end{figure}

We wanted to determine how the boundary conditions could influence the dynamics of the coupled brain-body-environment system. As we noted in our description of the mixed PDE-ODE system, the agent is coupled to the environment at a point along $x$, $x_e$. $x_e$ is the value from which $r$ is integrated over change in $v$. Intuitively, if the attractor metaphor for the boundary conditions was correct, then we would expect placing $x_e$ closer to the boundary would have the same effect as increasing $K$. To test this hypothesis, we replicated the previous plot but instead of varying $K$, we varied the position of $x_e$. We plotted the results in
diagram in \ref{fig:xe-diag}.

As we can see the approximation holds up relatively well. We can replicate the way in which $K$ shifts the phase boundaries. Again the effect of $A$ is mitigated. As $x_e$ gets closer to $x_max$, the effect of the boundary condition gets stronger. The tradeoff for the stronger boundary effect is a lower impact of the ongoing feedback dynamics. This results in a bias toward stability, even as $A$ increases the dynamics that propagate to $r$ remain stable. Thus the effects of $A$ on the overall system are significantly reduced.

Neurally, this process resembles the production of strong bias of a motor program. Placing $x_e$ in the middle of $x$ weakens the effect of the boundary condition. In that case, the dynamics at $x_e$ are dominated by the sensory and feedback dynamics between $v$ and $u$. As a result, $r$ is being driven by this sensory feedback and processing. In the alternative case when $x_e$ is close to $x_{max}$, the dynamics of $x_e$ are relatively free from these feedback dynamics and the behavior of $r$ is driven by whatever dynamic pattern exists at the boundary condition.

\section{Activity Gradients and Sloped Biases}
From the previous analysis, we found that the boundary condition could be understood as a neural bias which attracts parts of the system that are close to it. However, this also begs more question about the effects of the value of this neural bias on the dynamics. In particular, we considered the case where the two boundary conditions have different values. To explore this condition we set $u(x_{min},t)$ to be $0$ and varied the value of $u(x_{max},t)$. 

\begin{figure}
    \centering
    \includegraphics[width=0.8\linewidth]{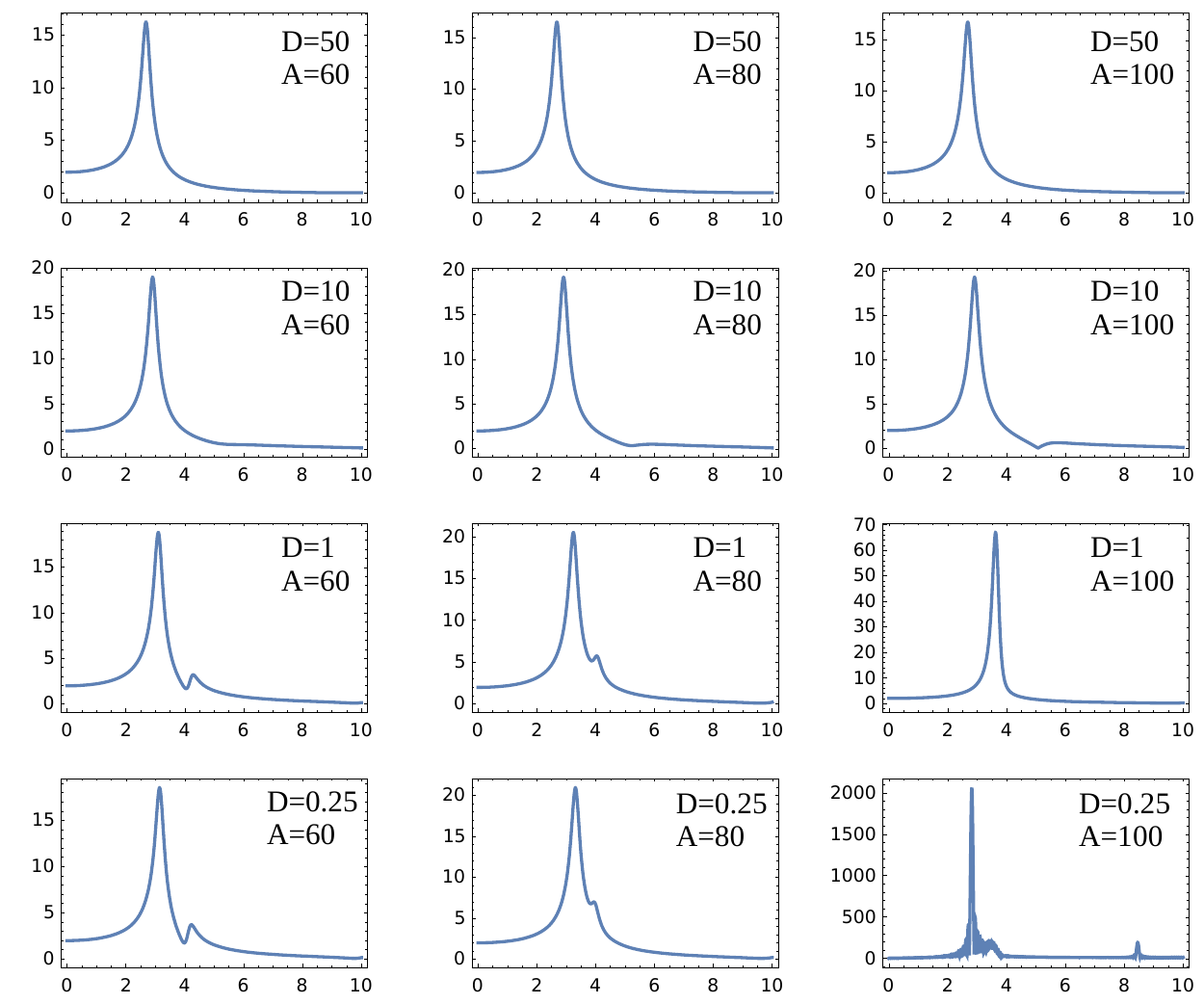}
    \caption{Frequency power spectrum of the angular velocity $\omega(t)$ as the bias slope $D$ and the sensory coupling strength $A$ are varied. The diffusivity is fixed at $K=1$ and the environment interacts globally in the agent $\sigma_s\rightarrow\infty$. The effector is at $x_e=0$, while the boundary condition is $BC(x)=D(x-x_{\sf min})$. We observe a similar variety of phases of the mixed PDE-ODE system but the transition point and the approach to criticality is modified due to the influence of the bias slope $D$ between the two ends of the agent.}
    \label{fig:slope-diag}
\end{figure}

We imagine that as we vary the value at $x_{max}$, we should see different effects emerge as a result of the feedback coupling between the agent and the environment. Note that as before, we set $x_e$ to be fixed in the middle so as to determine the contributions at the boundary conditions. We can visualize our manipulation as placing the string of our system along a slope. We refer to this slope as $D$. Based on our interpretation of the boundary condition as providing a bias, we assumed that the effect of increasing $D$ would be to shift the phase boundaries as specified previously. We plotted the dynamics in figure \ref{fig:slope-diag}. 

Looking at the resulting dynamics, we can see that the indeed that increasing $D$ does shift the relative boundaries of the phases. In particular we see that higher values of $D$ result in significantly longer ranges of the dampened regime even as we increase $A$. This happens because the slope enforces a biased value at $x_e$, this bias is what then dominates the activity of $r$. Since this activity is no longer correlated with the sensory input it does not lead to an oscillation in the environment variables and leads to the breadth of the dampened phase.

We can explain the dynamics in neural terms by considering $D$ as representing a neural activity gradient. The value of $D$ then determines the slope of this gradient. Such neural gradients create a bias in the neural dynamics where signals require high strength to propagate along the gradient.

\section{A Biased Path Between Sensory and Motor} 
For all of our analysis so far, we have kept a uniform distribution for $x_s$ over the field $v$. However, a more realistic approach would be to localize $x_s$ at a position. However, to maintain $v$ as a continuous field, we also need a convolution that would spread the effects of sensory coupling across $x$. We use a gaussian convolution centered at $x_s$ on the temporal dynamics of $v$ to model this aspect. We can see this diagrammatically in \ref{fig:agent-diagram}. Such an assumption is an oversimplification but allows to us to ask question regarding the distance between $x_s$ and $x_e$.

\begin{figure}
    \centering
    \includegraphics[width=1.1\linewidth]{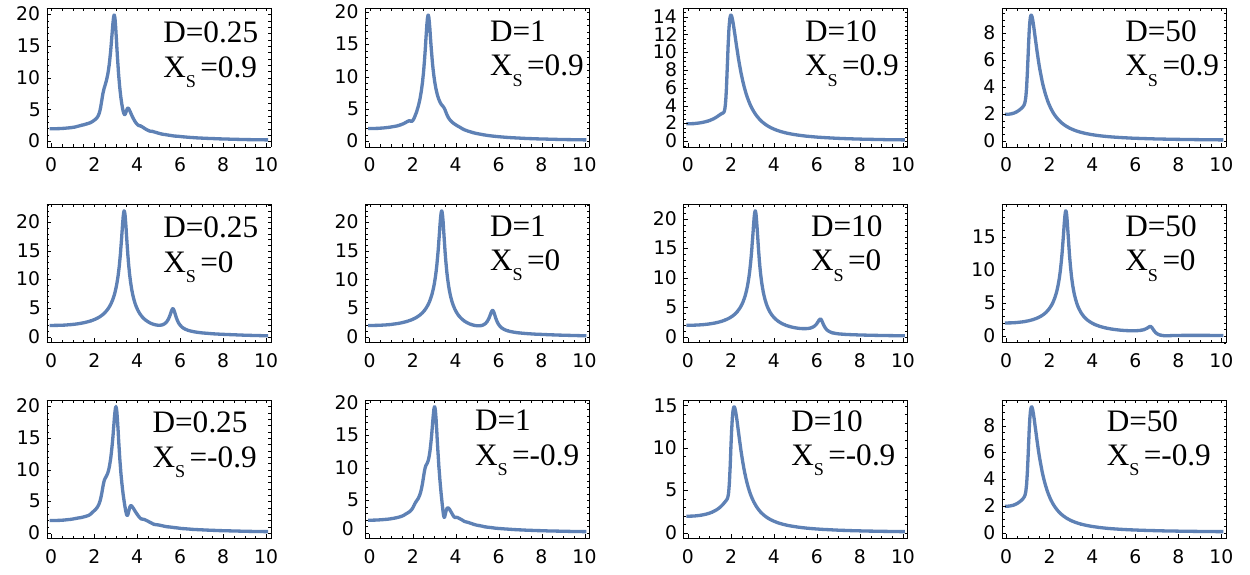}
    \caption{Frequency power spectrum of the angular velocity $\omega(t)$ as the bias slope $D$ and the sensory surface center $x_s$ are varied. The agent diffusivity is $K=1$, the activation sensitivity is $A=100$ and the environment interacts globally in the agent $\sigma_s=0.1$. The effector is at $x_e=0$, while the boundary condition is $BC(x)=D(x-x_{\sf min})$. We observe a similar variety of phases of the mixed PDE-ODE system but the transition point and the approach to criticality is modified due to the bias slope $D$ is topologically screened from interacting with the environment,specifically depending on where the sensory surface $x_s$ interacts with the agent.}
    \label{fig:dist-diag}
\end{figure}

Previously, we had established that the role of the variable $D$ was to introduce a slope of activity and that this slope represented a graident of neural bias. To understand the role of the distance between neural sensors and neural effectors, we varied the position of $x_s$ and the slope $D$, keeping $x_e$ fixed at the center. We plot this in firgure \ref{fig:dist-diag}.

Looking at the figure, we see that at the low slope the dominant frequencies are capable of making it from $x_s$ to $x_e$ even under very high bombardment of sensory input. The effects of increasing $D$ as before is to introduce a strong bias that constrains the dynamics of the system. However, the distance between $x_s$ and $x_e$ also creates a unique effect in that it enforces a specific pattern on the dynamics which in turn delays the onset of chaos.

Neurally, this corresponds to the case where between sensory input and motor effectors in the brain there is a range of neurons with graded activity. Our results show that such graded activity can cause constraints on signal propagation and might even be strong enough to select specific frequencies. We see this as an exciting avenue for future research.
\section{Results}

We began by identifying the dynamical regimes exhibited by the system, leveraging bifurcation theory to demonstrate how these dynamics depend critically on the sensory coupling parameter. Specifically, we showed that increasing the strength of sensory coupling drives the system through distinct transitions: from stable equilibrium dynamics, to oscillatory behavior, and eventually to chaotic dynamics. To contextualize these transitions, we linked them to bifurcation points (e.g., Hopf and period-doubling bifurcations) that mark shifts between regimes, though further numerical validation or analytical approaches could strengthen this claim since, here we relied on numerical branch following continuation methods.

Next, we explored the introduction of spatial terms to the brain ($u$) and body ($v$) variables. By incorporating spatial diffusion, we observed that the propagation of signals across $u$ and $v$ dampens the destabilizing effects of sensory input. This spatialization allows the system to tolerate significantly stronger sensory coupling without collapsing into fully chaotic dynamics. To unpack this phenomenon, we employed analysis of the boundary conditions that was informed by our understanding of the spatial dynamics. This simplification retained the core dynamical effects while reducing computational complexity, though future work could quantify the trade-offs of this approximation.

Finally, we investigated the role of sensor and effector positions $x_s,x_e$ along $v$. Our results revealed that the relative positioning of the effector and sensory surfaces critically shapes signal flow. Specifically, gradients in internal neural excitability between these regions mediate how freely sensory signals propagate to motor outputs, suggesting that morphological alignment (e.g., proximity or neural connectivity gradients) is key to functional sensorimotor integration.

\section{Discussion}
In the realm of brain-body-environment modeling, consideration of spatial attributes of agents and environments can enhance verisimilitude with respect to the observed reality. In this work, we provide a valuable example of how this approach enriches the resulting analysis in a way that emphasizes relevant dynamics by extending the model first proposed by \cite{ThornileyHusbands2013}.

Our findings contribute to growing work on brain-body-environment systems \citep{PfeiferBongard2007, Clark1997}, which emphasizes how spatial structure—whether in the environment or the agent’s physiology—constrains cognitive and behavioral possibilities. Prior studies have highlighted environmental spatiality, but here we demonstrate the inverse: the agent’s internal spatial structure (e.g., neural geometry, diffusion dynamics) fundamentally alters its capacity to process sensory information and stabilize behavior. For instance, as we have shown, spatial diffusion smooths chaotic fluctuations into coherent oscillations, exemplifying morphological computation—where physical structure itself performs computational work. This challenges traditional neural network frameworks that prioritize topology over geometry, raising new questions: How does neural layout (e.g., 1D vs. 2D) affect emergent dynamics? Could spatial gradients substitute for explicit learning in some tasks?

This work aligns with broader efforts to model agents as physically embedded systems. Brain-body-environment approaches \citep{ChielBeer1997} reject computationalist abstractions, instead emphasizing how materiality (e.g., body mechanics, environmental dynamics) co-constitutes cognition. Our model extends this view by treating the brain itself as a spatially extended medium, not just a control circuit. Just as limbs and environments interact through physics, neural tissue’s material properties (e.g., diffusion rates, asymmetry) shape its dynamics. Future work might test these insights in embodied robots or biological preparations.

\section{Acknowledgments}
We would like to thank Professor Gerardo Ortiz for access to computing cluster used for all numerical simulation in this work. We would also like to thank Doga Ozcan and Leigh Levinson for their support during the writing process. Finally, we would like to thank the Indiana Graduate Workers Coalition for bringing us together.

\footnotesize
\bibliographystyle{apalike}
\bibliography{example} 

\begin{thebibliography}{}

\bibitem[Beer, 1995]{Beer1995}
Beer, R.~D. (1995).
\newblock A dynamical systems perspective on agent-environment interaction.
\newblock {\em Artificial intelligence}, 72(1-2):173--215.

\bibitem[Beer, 2000]{Beer2000}
Beer, R.~D. (2000).
\newblock Dynamical approaches to cognitive science.
\newblock {\em Trends in Cognitive Sciences}, 4(3):91--99.

\bibitem[Beer, 2003]{Beer2003}
Beer, R.~D. (2003).
\newblock The dynamics of active categorical perception in an evolved model agent.
\newblock {\em Adaptive Behavior}, 11(4):209--243.

\bibitem[Bose and Ghosh, 2019]{bose_bifurcation_2019}
Bose, I. and Ghosh, S. (2019).
\newblock Bifurcation and criticality.
\newblock {\em Journal of Statistical Mechanics: Theory and Experiment}, 2019(4):043403.
\newblock Publisher: IOP Publishing and SISSA.

\bibitem[Bressloff, 2012]{Bressloff2012}
Bressloff, P.~C. (2012).
\newblock Spatiotemporal dynamics of continuum neural fields.
\newblock {\em Journal of Physics A: Mathematical and Theoretical}, 45(3):033001.

\bibitem[Chiel and Beer, 1997]{ChielBeer1997}
Chiel, H.~J. and Beer, R.~D. (1997).
\newblock The brain has a body: Adaptive behavior emerges from interactions of nervous system, body and environment.
\newblock {\em Trends in Neurosciences}, 20(12):553--557.

\bibitem[Clark, 1997]{Clark1997}
Clark, A. (1997).
\newblock {\em Being There: Putting Brain, Body, and World Together Again}.
\newblock MIT Press.

\bibitem[Coombes et~al., 2014]{CoombesBeimGrabenPotthastWright2014}
Coombes, S., beim Graben, P., Potthast, R., and Wright, J. (2014).
\newblock Neural fields: Theory and applications.
\newblock In {\em Springer Series in Computational Neuroscience}. Springer.

\bibitem[Dale and Husbands, 2010]{dale_evolution_2010}
Dale, K. and Husbands, P. (2010).
\newblock The {Evolution} of {Reaction}-{Diffusion} {Controllers} for {Minimally} {Cognitive} {Agents}.
\newblock {\em Artificial Life}, 16(1):1--19.

\bibitem[Faugeras et~al., 2009]{FaugerasEtAl2009}
Faugeras, O., Touboul, J., and Cessac, B. (2009).
\newblock A constructive mean-field analysis of multi-population neural networks with random synaptic weights and stochastic inputs.
\newblock {\em Frontiers in Computational Neuroscience}, 3:1.

\bibitem[Hutt and Rougier, 2010]{HuttRougier2010}
Hutt, A. and Rougier, N.~P. (2010).
\newblock Numerical simulation scheme of one- and two dimensional neural fields involving space-dependent delays.
\newblock {\em Neural Computation}, 22(8):2130--2163.

\bibitem[Kuznetsov, 2006]{Kuznetsov2006Hopf}
Kuznetsov, Y.~A. (2006).
\newblock Andronov-hopf bifurcation.
\newblock {\em Scholarpedia}, 1(10):1858.

\bibitem[Lange et~al., 2020]{lange_fourier_2020}
Lange, H., Brunton, S.~L., and Kutz, N. (2020).
\newblock From {Fourier} to {Koopman}: {Spectral} {Methods} for {Long}-term {Time} {Series} {Prediction}.
\newblock {\em arXiv:2004.00574 [cs, eess, stat]}.
\newblock arXiv: 2004.00574.

\bibitem[Marder, 2020]{marder_theoretical_2020}
Marder, E. (2020).
\newblock Theoretical musings.
\newblock {\em eLife}, 9:e60703.
\newblock Publisher: eLife Sciences Publications, Ltd.

\bibitem[Meiss, 2007]{Meiss2007PeriodDoubling}
Meiss, J. (2007).
\newblock Period doubling.
\newblock {\em Scholarpedia}, 2(2):1629.

\bibitem[Paik et~al., 2020]{PaikEtAl2020}
Paik, S.-B., Song, J.~H., Choi, W., and Lee, S.-H. (2020).
\newblock Amasine: Automated 3d mapping of single neurons for standardized analysis of brain-wide neural circuits.
\newblock {\em Cell Reports}, 31(8):107684.
\newblock As referenced in [6] for spatial organization of neural circuits.

\bibitem[Pfeifer and Bongard, 2007]{PfeiferBongard2007}
Pfeifer, R. and Bongard, J.~C. (2007).
\newblock {\em How the body shapes the way we think: A new view of intelligence}.
\newblock MIT Press.

\bibitem[Pinotsis et~al., 2013]{PinotsisEtAl2013}
Pinotsis, D.~A., Leite, M., and Friston, K.~J. (2013).
\newblock On conductance-based neural field models.
\newblock {\em Frontiers in Computational Neuroscience}, 7:158.

\bibitem[Risken, 1996]{Risken1996}
Risken, H. (1996).
\newblock The fokker-planck equation: Methods of solution and applications.
\newblock 18.

\bibitem[Strogatz, 2018]{Strogatz2018}
Strogatz, S.~H. (2018).
\newblock {\em Nonlinear Dynamics and Chaos: With Applications to Physics, Biology, Chemistry, and Engineering}.
\newblock CRC Press, 2nd edition.

\bibitem[Thorniley and Husbands, 2013]{ThornileyHusbands2013}
Thorniley, J. and Husbands, P. (2013).
\newblock Information hiding in embodied dynamical systems.
\newblock In {\em Proceedings of the European Conference on Artificial Life (ECAL 2013)}, pages 513--520. MIT Press.

\bibitem[Varela et~al., 1991]{VarelaThompsonRosch1991}
Varela, F.~J., Thompson, E., and Rosch, E. (1991).
\newblock {\em The embodied mind: Cognitive science and human experience}.
\newblock MIT Press.

\end{thebibliography}

\end{document}